\documentstyle[12pt,jeep]{article}

\catcode`\@=11

\let\DOTSI\relax
\def\RIfM@{\relax\ifmmode}
\def\FN@{\futurelet\next}
\newcount\intno@
\def\iint{\DOTSI\intno@\tw@\FN@\ints@}
\def\iiint{\DOTSI\intno@\thr@@\FN@\ints@}
\def\iiiint{\DOTSI\intno@4 \FN@\ints@}
\def\idotsint{\DOTSI\intno@\z@\FN@\ints@}
\def\ints@{\findlimits@\ints@@}
\newif\iflimtoken@
\newif\iflimits@
\def\findlimits@{\limtoken@true\ifx\next\limits\limits@true
 \else\ifx\next\nolimits\limits@false\else
 \limtoken@false\ifx\ilimits@\nolimits\limits@false\else
 \ifinner\limits@false\else\limits@true\fi\fi\fi\fi}
\def\multint@{\int\ifnum\intno@=\z@\intdots@                                %1
 \else\intkern@\fi                                                          %2
 \ifnum\intno@>\tw@\int\intkern@\fi                                         %3
 \ifnum\intno@>\thr@@\int\intkern@\fi                                       %4
 \int}                                                                      %5
\def\multintlimits@{\intop\ifnum\intno@=\z@\intdots@\else\intkern@\fi
 \ifnum\intno@>\tw@\intop\intkern@\fi
 \ifnum\intno@>\thr@@\intop\intkern@\fi\intop}
\def\intic@{\mathchoice{\hskip.5em}{\hskip.4em}{\hskip.4em}{\hskip.4em}}
\def\negintic@{\mathchoice
 {\hskip-.5em}{\hskip-.4em}{\hskip-.4em}{\hskip-.4em}}
\def\ints@@{\iflimtoken@                                                    %1
 \def\ints@@@{\iflimits@\negintic@\mathop{\intic@\multintlimits@}\limits    %2
  \else\multint@\nolimits\fi                                                %3
  \eat@}                                                                    %4
 \else                                                                      %5
 \def\ints@@@{\iflimits@\negintic@
  \mathop{\intic@\multintlimits@}\limits\else
  \multint@\nolimits\fi}\fi\ints@@@}
\def\intkern@{\mathchoice{\!\!\!}{\!\!}{\!\!}{\!\!}}
\def\plaincdots@{\mathinner{\cdotp\cdotp\cdotp}}
\def\intdots@{\mathchoice{\plaincdots@}
 {{\cdotp}\mkern1.5mu{\cdotp}\mkern1.5mu{\cdotp}}
 {{\cdotp}\mkern1mu{\cdotp}\mkern1mu{\cdotp}}
 {{\cdotp}\mkern1mu{\cdotp}\mkern1mu{\cdotp}}}

\newif\iffirstchoice@
\firstchoice@true
\def\textfonti{\the\textfont\@ne}
\def\textfontii{\the\textfont\tw@}
\def\text{\RIfM@\expandafter\text@\else\expandafter\text@@\fi}
\def\text@@#1{\leavevmode\hbox{#1}}
\def\text@#1{\mathchoice
 {\hbox{\everymath{\displaystyle}\def\textfonti{\the\textfont\@ne}%
  \def\textfontii{\the\textfont\tw@}\textdef@@ T#1}}
 {\hbox{\firstchoice@false
  \everymath{\textstyle}\def\textfonti{\the\textfont\@ne}%
  \def\textfontii{\the\textfont\tw@}\textdef@@ T#1}}
 {\hbox{\firstchoice@false
  \everymath{\scriptstyle}\def\textfonti{\the\scriptfont\@ne}%
  \def\textfontii{\the\scriptfont\tw@}\textdef@@ S\rm#1}}
 {\hbox{\firstchoice@false
  \everymath{\scriptscriptstyle}\def\textfonti
  {\the\scriptscriptfont\@ne}%
  \def\textfontii{\the\scriptscriptfont\tw@}\textdef@@ s\rm#1}}}
\def\textdef@@#1{\textdef@#1\rm\textdef@#1\bf\textdef@#1\sl\textdef@#1\it}
\def\DN@{\def\next@}
\def\eat@#1{}
\def\textdef@#1#2{%
 \DN@{\csname\expandafter\eat@\string#2fam\endcsname}%
 \if S#1\edef#2{\the\scriptfont\next@\relax}%
 \else\if s#1\edef#2{\the\scriptscriptfont\next@\relax}%
 \else\edef#2{\the\textfont\next@\relax}\fi\fi}

\def\Let@{\relax\iffalse{\fi\let\\=\cr\iffalse}\fi}
\def\vspace@{\def\vspace##1{\crcr\noalign{\vskip##1\relax}}}
\def\multilimits@{\bgroup\vspace@\Let@
 \baselineskip\fontdimen10 \scriptfont\tw@
 \advance\baselineskip\fontdimen12 \scriptfont\tw@
 \lineskip\thr@@\fontdimen8 \scriptfont\thr@@
 \lineskiplimit\lineskip
 \vbox\bgroup\ialign\bgroup\hfil$\m@th\scriptstyle{##}$\hfil\crcr}
\def\Sb{_\multilimits@}
\def\endSb{\crcr\egroup\egroup\egroup}
\def\Sp{^\multilimits@}

\newdimen\ex@
\def\rightarrowfill@#1{$#1\m@th\mathord-\mkern-6mu\cleaders
 \hbox{$#1\mkern-2mu\mathord-\mkern-2mu$}\hfill
 \mkern-6mu\mathord\rightarrow$}
\def\leftarrowfill@#1{$#1\m@th\mathord\leftarrow\mkern-6mu\cleaders
 \hbox{$#1\mkern-2mu\mathord-\mkern-2mu$}\hfill\mkern-6mu\mathord-$}
\def\leftrightarrowfill@#1{$#1\m@th\mathord\leftarrow\mkern-6mu\cleaders
 \hbox{$#1\mkern-2mu\mathord-\mkern-2mu$}\hfill
 \mkern-6mu\mathord\rightarrow$}
\def\overrightarrow{\mathpalette\overrightarrow@}
\def\overrightarrow@#1#2{\vbox{\ialign{##\crcr\rightarrowfill@#1\crcr
 \noalign{\kern-\ex@\nointerlineskip}$\m@th\hfil#1#2\hfil$\crcr}}}

\def\overleftarrow{\mathpalette\overleftarrow@}
\def\overleftarrow@#1#2{\vbox{\ialign{##\crcr\leftarrowfill@#1\crcr
 \noalign{\kern-\ex@\nointerlineskip}$\m@th\hfil#1#2\hfil$\crcr}}}
\def\overleftrightarrow{\mathpalette\overleftrightarrow@}
\def\overleftrightarrow@#1#2{\vbox{\ialign{##\crcr\leftrightarrowfill@#1\crcr
 \noalign{\kern-\ex@\nointerlineskip}$\m@th\hfil#1#2\hfil$\crcr}}}
\def\underrightarrow{\mathpalette\underrightarrow@}
\def\underrightarrow@#1#2{\vtop{\ialign{##\crcr$\m@th\hfil#1#2\hfil$\crcr
 \noalign{\nointerlineskip}\rightarrowfill@#1\crcr}}}

\def\underleftarrow{\mathpalette\underleftarrow@}
\def\underleftarrow@#1#2{\vtop{\ialign{##\crcr$\m@th\hfil#1#2\hfil$\crcr
 \noalign{\nointerlineskip}\leftarrowfill@#1\crcr}}}
\def\underleftrightarrow{\mathpalette\underleftrightarrow@}
\def\underleftrightarrow@#1#2{\vtop{\ialign{##\crcr$\m@th\hfil#1#2\hfil$\crcr
 \noalign{\nointerlineskip}\leftrightarrowfill@#1\crcr}}}

\catcode`\@=\active
\def\frac#1#2{{#1 \over #2}}

\newcount\GRAPHICSTYPE
\def\GRAPHICSPS#1{%
\ifnum\GRAPHICSTYPE=1 language "PS", include "#1"\else%
ps: #1\fi}

\def\graffile#1#2#3#4{\leavevmode\raise -#4 \hbox{%
\raise #3 \hbox{\rule{0.003in}{0.003in}\special{#1}}}%
{\raise -#4 \hbox to #2 {\vrule height#3 width0in depth0in\hfil}}%
}

\def\draftbox#1#2#3#4{\leavevmode\raise -#4 \hbox{\frame{\rlap{\protect\tiny
#1}%
\hbox to #2{\vrule height#3 width0in depth0in\hfil}}}}

\newcount\draft
\def\GRAPHIC#1#2#3#4#5{\ifnum\draft=1 \draftbox{#2}{#3}{#4}{#5}\else%
\graffile{#1}{#3}{#4}{#5}\fi}

\def\addtoLaTeXparams#1{\edef\LaTeXparams{\LaTeXparams #1}}

\def\doFRAMEparams#1{\readFRAMEparams#1\end}
\def\readFRAMEparams#1{%
\ifx#1\end%
\let\next=\relax%
\else%
\ifx#1i%
\dispkind=0%
\fi%
\ifx#1d%
\dispkind=1%
\fi%
\ifx#1f%
\dispkind=2%
\fi%
\ifx#1t%
\addtoLaTeXparams{t}%
\fi%
\ifx#1b%
\addtoLaTeXparams{b}%
\fi%
\ifx#1p%
\addtoLaTeXparams{p}%
\fi%
\ifx#1h%
\addtoLaTeXparams{h}%
\fi%
\let\next=\readFRAMEparams%
\fi%
\next%
}

\def\IFRAME#1#2#3#4#5{\GRAPHIC{#5}{#4}{#1}{#2}{#3}}

\def\DFRAME#1#2#3#4{
  \begin{center}
    \GRAPHIC{#4}{#3}{#1}{#2}{0in}
  \end{center}
}

\def\FFRAME#1#2#3#4#5#6#7{
  \begin{figure}[#1]
    \begin{center}
      \GRAPHIC{#7}{#6}{#2}{#3}{0in}
    \end{center}
    \caption{\label{#5}#4}
  \end{figure}
}

\def\FRAME#1#2#3#4#5#6#7#8{%
\newcount\dispkind%
\def\LaTeXparams{}%
\dispkind=0%
\def\LaTeXparams{}%
\doFRAMEparams{#1}%
\ifnum\dispkind=0%
\IFRAME{#2}{#3}{#4}{#7}{#8}%
\else
  \ifnum\dispkind=1
    \DFRAME{#2}{#3}{#7}{#8}
  \else
    \ifnum\dispkind=2
      \FFRAME{\LaTeXparams}{#2}{#3}{#5}{#6}{#7}{#8}
    \fi
  \fi
\fi
}

\catcode`\@=11
\long\def\QQQ#1#2{}
\def\QTP#1{}
\long\def\QQA#1#2{}

\def\EXPAND#1[#2]#3{}
\def\NOEXPAND#1[#2]#3{}

\def\LaTeXparent#1{}

\@ifundefined{INDEX}{\def\INDEX#1#2{}{}}{}
\@ifundefined{SUBINDEX}{\def\SUBINDEX#1#2#3{}{}{}}{}
\def\initial#1{\bigbreak{\raggedright\large\bf #1}\kern 2pt\penalty3000}

\@ifundefined{abstract}{%
\def\abstract{\if@twocolumn
\section*{Abstract (Not appropriate in this style!)}
\else \small
\begin{center}
{\bf Abstract\vspace{-.5em}\vspace{0pt}}
\end{center}
\quotation
\fi}}{}
\@ifundefined{endabstract}{%
\def\endabstract{\if@twocolumn\else\endquotation\fi}}{}
\@ifundefined{maketitle}{\def\maketitle#1{}}{}
\@ifundefined{affiliation}{\def\affiliation#1{}}{}
\@ifundefined{proof}{}{}
\@ifundefined{newfield}{\def\newfield#1#2{}}{}
\@ifundefined{chapter}{\def\chapter#1{\par(Chapter head:)#1\par }}{}
\@ifundefined{part}{\def\part#1{\par(Part head:)#1\par }}{}
\@ifundefined{section}{\def\part#1{\par(Section head:)#1\par }}{}
\@ifundefined{subsection}{\def\part#1{\par(Subsection head:)#1\par }}{}
\@ifundefined{subsubsection}{\def\part#1{\par(Subsubsection head:)#1\par }}{}
\@ifundefined{paragraph}{\def\part#1{\par(Subsubsubsection head:)#1\par }}{}
\@ifundefined{subparagraph}{\def\part#1{\par(Subsubsubsubsection head:)#1\par
}}{}

\newdimen\theight
\def \Column{%
             \vadjust{\setbox0=\hbox{\scriptsize\quad\quad tcol}%
             \theight=\ht0
             \advance\theight by \dp0    \advance\theight by \lineskip
             \kern -\theight \vbox to \theight{\rightline{\rlap{\box0}}%
             \vss}%
             }}%

\def\qed{\ifhmode\unskip\nobreak\fi\ifmmode\ifinner\else\hskip5\p@\fi\fi
 \hbox{\hskip5\p@\vrule width4\p@ height6\p@ depth1.5\p@\hskip\p@}}
\catcode`@=12 % at signs are no longer letters

\makeatletter

\begin{document}

%%%%%%%%%%%%%%%%%%%%%%%%%%%%%%%%%%%%

%\begin{titlepage}
\noindent {hep-th/yymmxxx}       \hfill                  {USC-93/HEP-B3}\\
\vspace{0.75cm}

\begin{centering}

{\huge QCD AND STRINGS IN 2D\footnote {Based on lecture delivered at the
Strings '93 Conference, Berkeley, CA, May 1993. }}\\
\vspace{1cm}
{\large Itzhak Bars\footnote{Research supported in part by the
DOE Grant No. DE-FG03-84ER-40168.}\\
Department of Physics and Astronomy\\
University of Southern California\\
Los Angeles, CA 90089-0484, USA}\\
\end{centering}
\vspace{.25cm}

\begin{abstract}
\vspace {.2cm}
In two dimensions large N QCD with quarks, defined on the plane, is
equivalent to a modified string theory with quarks at the ends and taken in
the zero fold sector. The equivalence that was established in 1975 was
expressed in the form of an interacting string action that reproduces the
spectrum and the 1/N interactions of 2D QCD. This action may be a starting
point for an analytic continuation to a four dimensional string version of
QCD. After reviewing the old work I discuss relations to recent developments
in the pure QCD-string equivalence on more complicated background geometries.
\end{abstract}

\section{Introduction}

During the past year, starting with the work of Gross \cite{Gross}, there
has been new progress in establishing a definite relationship between
large-N QCD and strings in 2-dimensions \cite{Minahan,GoTa,MiPo,Doug,DoKa}.
Pure QCD (without quarks) is trivial in 2 dimensions if it is defined on the
plane. However, when defined on more complicated geometries, such as the
cylinder, torus, pretzel, etc. there remains non-contractable Wilson lines
(color glue strings) that survive as degrees of freedom. They have shown, to
all orders in the 1/N expansion, that the path integral for large N QCD can
be reorganized as a sum over surfaces analogous to the Polyakov path
integral over all Riemann surfaces in string theory (all string loops). The
parameter 1/N plays the role of the string coupling constant. This shows
that there must be a string action from which this string path integral may
be derived. However, so far it has not been possible to identify this
QCD-string action.

On the other hand, back in 1975 we established other definite quantitative
relationships between large-N QCD and strings \cite{BBHP,BH,BAA,BAB}. In
that case quarks were included, but QCD was defined on the plane. Therefore,
the new results are complementary to the old ones. Furthermore, in contrast
to the recent progress, the old correspondance established a many-body
formulation of an {\it interacting} {\it string action} \cite{BAA,BAB} that
reproduced all the features of QCD on the plane, including the large-N
propagators, and the $1/N$ interaction vertices. In explicit computations to
first order in $1/N$ this string action gave the same results as QCD for the
spectrum, the 3-meson form factors and the electroweak form factors. For
agreement with QCD the string had to be limited to one topological sector,
namely to the no-fold sector. This sector was self contained and consistent
within the string theory. The interacting Hamiltonian that was derived from
the string action implemented the restriction to the no fold sector
consistently \cite{BAB}. The recent developments have also emphasized the no
fold sector.

The string theory version allowed additional topological sectors in which
the string folded on itself. The folded string was analysed both classically
and quantum mechanically \cite{BBHP}. The semi-classical spectrum of the
folded string which was derived in \cite{BBHP} was later reproduced by the
semi-classical spectrum of the Liouville mode discovered by Polyakov \cite
{POL}. Thus, there is a close correspondance between the degrees of freedom
provided by the folds on the one hand and the degrees of freedom of the
Liouville field on the other hand. One may therefore associate the folds
with the dynamics that relate to 2D gravity. The fold degrees of freedom
were obviously absent in 2D QCD on the plane. On the other hand, from
another point of view the folds could also be interpreted as points carrying
color charge (i.e. dynamical degree of freedom, such as gluons or other
adjoint matter), while the strings in between could be interpreted as the
glue on a Wilson line \cite{BAA}
{\footnote{This point of view, first suggested in Ref.\cite{BAA}, seems to have
found real applications in
four dimensions. Our model was later adopted by  several phenomenological
groups that applied the 2D string dynamics to the fragmentation
of hadrons and hadronization of quark and
gluon jets in various high energy reactions in four dimensions.
It is my understanding that the best
phenomenological fit to  hadronization or fragmentation is a model
that utilizes our 2D
string theory (with folds representing gluons) for the transverse
dynamics of quarks and gluons. The model is known to phenomenologists as
the ``Lund Model" \cite {Andersson}.
The 2D string theory with no folds has also been useful in
the discussion of other aspects of hadrons in four dimensions \cite{PISARSKI}.}
}. These were also obviously absent in 2D\ QCD\ on the plane. But in fact,
this interpretation of folds is realized in a hybrid lattice formulation of
4D QCD later developed by Bardeen and Pierson \cite{BaPi}. In trying to make
a connection to the more recent developments that involve more complicated
geometries or more complicated versions of QCD, there seems to be room for
the folds that fit either kind of interpretation.

In this lecture I will first briefly review the old QCD-string
correspondance on the plane, which includes quarks. In particular I will
emphasize the action which may be analytically continued to four dimensions.
Other lecturers in this conference will review the very interesting recent
progress for pure QCD on more complicated fixed (non-dynamical) geometries,
but without quarks or an action. In the second part of my lecture I will
discuss the folds, and in the last part of my lecture I will speculate on
possible relations between the new results and the old ones in a way that
considers the folds.

\section{Interacting action of strings and quarks and \hfill\break
equivalence to 2D QCD on the plane}

In 1975 'tHooft \cite{'tHooft} derived an integral equation that determines
the spectrum of large-N QCD. Soon afterwards Bardeen, Bars, Hanson and
Peccei \cite{BBHP}, working in 2D string theory, showed that the Nambu
string modified {\it with dynamical points added to the ends}, and taken in
the {\it zero fold sector} has the same spectrum as large-N QCD
{\footnote{It is important to emphasize that the 'tHooft equation for bosonic
quarks
(i.e. scalars in the fundamental representation) is different than the one for
fermions,
and it does not match any known string theory.}}. The zero fold sector is a
self consistent sector of the string+points theory. This sector has a
massless ``pion'' (not a tachyon) in the zero quark mass limit \cite
{'tHooft,Pak}. Bars and Hanson \cite{BH} modified the action at the
end points by adding quark degrees of freedom that carry spin and
color/flavor charges and showed how the string+quark system can interact
with electroweak gauge bosons. Furthermore, Bars \cite{BAA,BAB} formulated
an interacting string action in a many body formalism, by adding a term to
the action which fused the end point quarks on different strings, or created
a pair of quarks by splitting a string. The fusing/splitting occurs when the
end points on different strings touch each other, and involves the spins of
the quarks in a non-trivial manner. Making all terms gauge invariant with
respect to the electroweak interactions gave a further modification of the
total action. This final form of the action is

\begin{equation}
\label{action}S_{total}=\sum_{mesons}S_{meson}+
\sum_{I,J}S_{fuse/split}(I,J)+\sum_{I,J}S_{electroweak}(I,J)
\end{equation}
The action for a single meson is

\begin{equation}
\label{meson}S_{meson}=\int d\tau L_0(x_1(\tau ))+\int d\tau L_0(x_2(\tau
))+\gamma _M\int d\tau \int_0^\pi d\sigma \sqrt{-g}
\end{equation}
where the string $x^\mu (\tau ,\sigma )$ has end points $x_1^\mu (\tau
)=x^\mu (\tau ,\sigma =0)$ and $x_2^\mu (\tau )=x^\mu (\tau ,\sigma =\pi )$
. The last term is the Nambu action, while the first two terms are the world
line actions for the quarks at the end points

\begin{equation}
\label{wline}L_0(x_I(\tau ))=\overline{\psi }_I(\tau )i\overleftrightarrow{
\partial _\tau }\gamma _\mu \psi _I(\tau )\frac{x_{\tau I}^\mu (\tau )}{2
\sqrt{-x_{\tau I}^2}}-\sqrt{-x_{\tau I}^2(\tau )}\,\overline{\psi }_I(\tau
)\,m\,\psi _I(\tau )
\end{equation}
where $x_{\tau I}^\mu (\tau )=\partial x_I^\mu (\tau )/\partial \tau $ is
the velocity vector of the end point labelled by $I$, and $\psi _{\alpha
I}^{ia}(\tau )$ is the Dirac wavefunction for the quark attached to the same
end point, with its spin ``$\alpha ",$ color ``$i"$
{\footnote{The color index is used only as a counting device in the 2D string
theory.
It is assumed that there are only color singlet states. Factors of 1/N
eventually
appear because of the color counting. }} and flavor ``$a$'' labels.
Furthermore, $m_{ab}=m_a\delta _{ab}$ is a mass matrix in flavor space. The
'tHooft spectrum for mesons, in the leading term of the $1/N$ expansion of
QCD, is identically reproduced by the string meson action $S_{meson}$ in the
zero fold sector (which is a self consistent sector even after interactions
are included).

The index $I$ runs over all end points on all the strings that represent the
different mesons in the many body formalism. The interactions for joining
and/or splitting of any two end points is

\begin{equation}
\label{int}
\begin{array}{c}
{\sum }_{I,J}S_{join/split}(I,J)=-(f/4){\sum }_{I\neq J}
\int d\tau d\tau ^{\prime }\,\delta \left( x_I^\mu (\tau )-x_J^\mu (\tau
^{\prime })\right) \;\sqrt{-x_{\tau I}^2(\tau )}\;\sqrt{-x_{\tau J}^2(\tau
^{\prime })} \\ \qquad \ \quad \times \{\overline{\psi }_I(\tau )\left( i
\overleftarrow{\partial _\tau }\frac{\gamma \cdot x_{\tau I}}{x_{\tau I}^2}
+m\right) \psi _J(\tau ^{\prime })+\overline{\psi }_J(\tau ^{\prime })\left(
i\frac{\gamma \cdot x_{\tau I}}{x_{\tau I}^2}\overrightarrow{\partial _\tau }
+m\right) \psi _I(\tau )\}
\end{array}
\end{equation}
where the delta function insures that the points are at the same location,
while the spins of the quarks $(I,J)$\ are also fused. The factors $\sqrt{
-x_{\tau I}^2(\tau )}\;$insure $\tau $ and $\tau ^{\prime }$
reparametrization invariance. The operator sandwiched between the $(I,J)$
quark wavefunctions is the Dirac operator in which the derivative is taken
along the tangent to the worldline (or velocity of the end point).The
equation of motion that follows from \ref{wline} also involves the same
operator. Therefore, if the string coupling for mesons $\gamma _M$ in \ref
{meson} vanishes, the quarks can perform only free motion and do not
interact. Thus, after accounting for the quark equations of motion, it is
evident that the interaction above can take place only in the presence of
strings. Indeed, the interaction Hamiltonian derived from \ref{int} in \cite
{BAB} turns out to be proportional to $\gamma _M$ . Furthermore, the overall
coefficient in \ref{int} is fixed to $f=\pi /2$ by crossing symmetry. This
interaction turns out to reproduce exactly the $1/N$ corrections in 2D\ QCD.
This was shown by comparing the detailed QCD\ results \cite{CCG} and string
results \cite{BAB} for the 3-meson vertex, which is a form factor describing
the decay of one meson to two mesons.

Electroweak interactions of the quarks can be introduced by replacing
ordinary derivatives by gauge covariant derivatives in the actions above, as
follows

\begin{equation}
\label{subs}\frac{x_{\tau I}^\mu (\tau )}{x_{\tau I}^2}i\partial _\tau
\longrightarrow \frac{x_{\tau I}^\mu (\tau )}{x_{\tau I}^2}i\partial _\tau
-eA^\mu (x_I(\tau ))\
\end{equation}
where the gauge field $A$ couples to the flavor indices. This produces the
electroweak interactions

\begin{equation}
\label{eweak}
\begin{array}{c}
\sum_{I,J}S_{electroweak}=\sum_I\int d\tau \,
\sqrt{-x_{\tau I}^2(\tau )}\,\overline{\psi }_I(\tau )\ \,e\gamma \cdot
A(x_I(\tau ))\,\ \psi _I(\tau ) \\ \qquad \qquad +(f/4)\sum_{I\neq J}\int
d\tau \int d\tau ^{\prime }\,\delta \left( x_I^\mu (\tau )-x_J^\mu (\tau
^{\prime })\right) \;
\sqrt{-x_{\tau I}^2(\tau )}\;\sqrt{-x_{\tau J}^2(\tau ^{\prime })} \\ \qquad
\ \quad \qquad \times e\{\overline{\psi }_I(\tau )\ A(x_I(\tau ))\ \psi
_J(\tau ^{\prime })+\overline{\psi }_J(\tau ^{\prime })\ A(x_I(\tau ))\ \psi
_I(\tau )\}
\end{array}
\end{equation}
where the first term comes from covariantizing \ref{wline} and the second
term comes from covariantizing \ref{int}. A smooth parametrization in $\tau $
or $\tau ^{\prime }$ require that each worldline be timelike. Then, the
first term in this expression describes the interaction of the electroweak
field with a timelike quark line, while the second term corresponds to the
creation or annihilation of a pair, where each member of the pair has a
timelike propagation. Therefore these two terms must be crossing symmetric
to each other. After computing some diagrams in momentum space, it was found
that crossing symmetry requires $f=\pi /2$ ; so gauge invariance plus
crossing symmetry fixes the overall coefficient of the interaction term in
\ref{int}. Computations in both QCD \cite{CCG,EINH} and string theory \cite
{BAA,BAB} have shown that the electroweak form factors of mesons agree in
detail in both theories.

Thus the string action \ref{action}, which is a modification of the Nambu
action, taken in the zero fold sector, agrees exactly with QCD on the plane,
including the $1/N$ corrections. The zero fold condition is implemented in
the derivation of the Hamiltonian \cite{BAB}. The spectrum is given by \ref
{meson} and the 1/N interaction vertices come from (\ref{int},\ref{eweak}).
Furthermore, the perturbative expansion of the total Hamiltonian has
diagrams that are in one to one correspondance with the quark loops in QCD.
Since the propagator (i.e. spectrum), the vertex, and the set of diagrams
are the same in both theories, one expects agreement to all orders of $1/N$
for all quark loops, without introducing gluon loops; however, this latter
point has not been checked in detail since the computations are not
available in either QCD\ or string theory.

Therefore, according to the old results, we have an action formulation for
the correspondance between string theory with quarks at the ends {\it on the
plane, }and QCD with quarks\ {\it on the plane.} Note that the string action
is written in any number of dimensions in flat spacetime, although it was
worked out in detail in only two dimensions. The question naturally arises
whether this string action may be valid for the correspondance between
strings and QCD in higher dimensions? Of course, the whole point of this
kind of exercise in two dimensions is based on the hope that a QCD-string
formulation that is valid in two dimensions might be ``analytically
continued'' to four dimensions and be used as a good approximation to
understand hadronic physics. We know that the form of string theory given
above is incomplete for all applications in four dimensions since closed
strings that represent glueballs and their interactions are not yet
included. However, one can still attempt a comparison of 4D\ QCD\ to the
above string action, by restricting oneself to only planar graphs of QCD
which do not include glue balls. With this in mind, I believe it is a good
exercise for someone to attempt a comparison in four dimensions of the
string theory defined above to {\it planar }QCD (but with $1/N$ corrections
due to quark loops, which make holes on the sheet, included in the
comparison).

If such a limited comparison in 4D is successful to begin with, then the
following additional suggested modifications of the string action may
reproduce the gluonic aspects of QCD: (1) Closed strings representing
glueballs should be included along with the meson action \ref{meson}, (2)
The interaction \ref{int} allows open strings to make transitions to closed
string, (3) An interaction that allows the string to interact at interior
points should be included. Such a many body type action formulation, to be
used at low orders of $1/N,$ could be used for phenomenological applications
to low energy hadronic physics since it would account for confinement and
Regge behaviour, etc.. A more ambitious step would be to find a field
theoretic string action instead of a many body type string action, but that
formalism would be harder to work with for computations of standard
processes.

\section{Folds in 2D string theory and longitudinal modes}

In \cite{BBHP} folded 2D strings were discussed using several approaches. It
was shown that the classical solutions for folded strings emerged in
different gauges and the gauge invariant motions were displayed graphically.
The quantum theory of folded open massless strings was solved
semi-classically in a formalism of action-angle variables (closed strings
could be done just easily, see below). The normal modes were identified and
shown to be equivalent to those of an additional longitudinal string degree
of freedom, and the Lorentz invariance of the system was proven in the
second paper in \cite{BBHP}. The {\it semi-classical }mass operator was
expressed as $\alpha ^{\prime }(M^2-m_0^2)=\sum_{n=1}^\infty \alpha
_{-n}\alpha _n$ with $m_0^2=const.$. Notice that the $n=0$ term is excluded
\cite{BBHP}. It has the spectrum $\alpha ^{\prime }(M^2-m_0^2)=\sum_nnl_n$
with $l_n=0,1,2,\cdot \cdot \cdot $ . The classical solutions showed that
each normal mode corresponds to an open string that has $n-1$ folds, wraps
around by folding upon itself in equal lengths, and oscillates. Later, with
Polyakov's discovery of the Liouville mode, one could associate
semi-classically the Liouville degrees of freedom with those of folded
strings.

The classical solutions for folded strings are most easily given in the
conformal gauge (for the same physics in other gauges see \cite{BBHP}) . The
classical equations of motions and constraints for a free string in 2D are
\begin{equation}
\label{eqs}\partial _{+}\partial _{-}x^\mu =0,\qquad (\partial _{+}x^\mu
)^2=0=(\partial _{-}x^\mu )^2.
\end{equation}
The solutions are $x^\mu (\tau ,\sigma )=x_L^\mu (\sigma ^{+})+x_R^\mu
(\sigma ^{-})$ , where the left and right movers depend on $\sigma ^{\pm
}=\frac 12(\tau \pm \sigma )$ respectively, and they are constrained by $
\partial x_L^0=\pm \partial x_L^1$ , $\partial x_R^0=\pm \partial x_R^1$ .
The $\pm $ can generally switch signs discontinuously in various regions of $
\sigma ^{\pm }$ , but the $x^\mu $'s have to be continuous to obtain a
continuous string. For a physical description, the time coordinate $x^0(\tau
,\sigma )=x_L^0(\sigma ^{+})+x_R^0(\sigma ^{-})$ must be monotonically
increasing with $\tau $ , while both $x^\mu (\tau ,\sigma )$ must be
periodic in $\sigma $ (this periodicity is obvious for closed strings, but
is also required for open strings to satisfy boundary conditions). Using the
remaining gauge invariance of conformal transformations one can fix the
gauge as $x_L^0=\frac{\alpha ^{\prime }E}{2\pi }(\tau +\sigma )/2\ ,\ x_R^0=
\frac{\alpha ^{\prime }E}{2\pi }(\tau -\sigma )/2\ $, so that $x^0=\frac{
\alpha ^{\prime }E}\pi \tau $ satisfies the periodicity and monotonicity
conditions. The other string coordinate may be written as
\begin{equation}
\label{sol}x^1(\tau ,\sigma )=q+\frac{\alpha ^{\prime }E}{2\pi }\,(f(\tau
+\sigma )+g(\tau -\sigma )),\qquad f^{\prime }(\tau +\sigma )=\pm 1,\quad
g^{\prime }(\tau -\sigma )=\pm 1.
\end{equation}
Thus, the functions $f,g$ see-saw with slopes $\pm 1$ in the range $[-\pi
,\pi ]$ and are periodic with a period of $2\pi $ . For the closed string
the two functions are independent, but for the open string $f=g$ . The
simplest example of such a function is $f=|\tau +\sigma |_{per}$ , i.e. the
absolute value taken in the range $[-\pi ,\pi ]$ and then periodically
repeated (see \cite{BBHP,BH} for pictures) . For the open string one gets
the solution
\begin{equation}
\label{exsol}x^1(\tau ,\sigma )=q+\frac{\alpha ^{\prime }E}{2\pi }(|\tau
+\sigma |_{per}+|\tau -\sigma |_{per})
\end{equation}
This function describes an open string stretched out without any folds, the
two ends oscillate against each other while the intermediate points move in
such a way as to allow the whole string to twist around when the two end
points pass each other (remember that $\sigma =[0,\pi ]$ for open strings).
The parameter $E$ is the energy of the system. For a closed string with $
\sigma =[-\pi ,\pi ]$ , taking for illustration $f=g$ , one obtains a closed
string looped around once, each point moving such that the loop twists while
the two end points (folds) oscillate against each other. At the end of one
period $\tau \rightarrow \tau +2\pi $ the string comes back to its original
shape. There are an infinite number of such periodic see-saw functions which
may be distinguished by the locations of the see-saw points in the range $
[-\pi ,\pi ]$ . The general folded string may fold and unfold during
different time intervals in one period $\tau =[0,2\pi ]$. However, the
normal modes correspond to strings that remain completely folded in equal
lengths and oscillate in that form \cite{BBHP}. Although we have referred to
such strings as ``folded'', it is hard to distinguish between strings that
fold or strings that wrap in the present formalism, since the string lies in
one space dimension. The physical content of these two cases is quite
different from the point of view of oriented color flux, and this point will
be addressed below.

To quantize the theory one may identify physical degrees of freedom by
fixing the gauge completely. This was done in \cite{BBHP} in a gauge that
differs from the conformal gauge above. Starting with the generally
covariant formalism, and using the $\tau $$,\sigma $ reparametrization
invariance, one may choose arbitrarily two functions. Thus, one may choose
(i) $x^0(\tau ,\sigma )=\frac{\alpha ^{\prime }E}{2\pi }\tau $ and (ii) $
\partial _\sigma x^1(\tau ,\sigma )=$independent of $\sigma $ in between $Z$
folds at $\sigma =\sigma _i,\ i=1,2,\cdot \cdot \cdot ,Z$ (i.e. $x^1(\tau
,\sigma )\;$ is linear in $\sigma $ in between the folds) $.$ This leaves
only the end points ($x^1(\tau ,0)\equiv x_0(\tau ),\ x^1(\tau ,\pi )\equiv
x_{Z+1}(\tau ))$ and the folds at $x^1(\tau ,\sigma _i)\equiv x_i(\tau )$
and their momenta $p_i$ as canonical degrees of freedom (note that at a fold
or end point $\partial _\sigma x^\mu =0$ and therefore these points must
move with the speed of light to satisfy the constraint $(\partial _\tau
x^\mu )^2=(\partial _\sigma x^\mu )^2)$. It was shown that, in the $Z$-fold
sector, the open string Hamiltonian is given by
\begin{equation}
\label{ham0}H=\sum_{i=0}^{Z+1}|p_i|+\frac 1{\alpha ^{\prime
}}\sum_{i=0}^Z|x_i-x_{i+1}|\ .
\end{equation}
The classical solutions of this Hamiltonian coincide with those displayed
above in the conformal gauge \cite{BBHP}. For a closed string with $2Z$
folds, one gets the same form, with a small modification that replaces the
lower limit by $i=1$, the upper limit on both sums by $2Z$ instead of $Z+1$
and $Z$ respectively, and also requires $x_{2Z+1}=x_1$ (redefining $x^1(\tau
,0)\equiv x_1(\tau )$ ) .

One may also discuss the theory in the lightcone gauge, $x^{+}(\tau ,\sigma
)=\frac{\alpha ^{\prime }P^{+}}\pi \tau $ , provided one slows down the
motion of the folds or end-points by making them massive, and then taking
the mass to zero at the end of the calculation (this is necessary because of
the singular lightcone description of massless particles that move with the
speed of light). The canonical lightcone degrees of freedom of the folds or
end-points are $(x_i^{-},p_i^{+}).$ The lightcone gauge open string
Hamiltonian takes the form
\begin{equation}
\label{ham+}P^{-}=\sum_{i=0}^{Z+1}\frac{m_i^2}{2p_i^{+}}+\frac 1{\alpha
^{\prime }}\sum_{i=0}^Z|x_i^{-}-x_{i+1}^{-}|\ .
\end{equation}
Again, the classical motions coincide with those of the conformal gauge {\it
in the limit} of zero masses $m_i\rightarrow 0$ \cite{BBHP}. The advantage
of the lightcone formalism is its obvious Lorentz covariance, but it can be
argued that in a certain sense the timelike gauge is Lorentz covariant as
well \cite{BBHP}. The closed string Hamiltonian is obtained with the same
modifications mentioned in the previous paragraph is
\begin{equation}
\label{hamm}P^{-}=\sum_{i=1}^{2Z}\frac{m_i^2}{2p_i^{+}}+\frac 1{\alpha
^{\prime }}\sum_{i=1}^{2Z}|x_i^{-}-x_{i+1}^{-}|\ .
\end{equation}
with $x_{2Z+1}^{-}=x_1^{-}$ .

The semi-classical quantization was carried out by transforming to
action-angle variables $(\vartheta _i,J_i)$ , and showing that the mass
operator takes the form $\alpha ^{\prime }$$M^2=\sum_iJ_i$. It was shown
that the $i$'th mode describes a string that is wrapped around $i$ times,
and that it covers a basic phase space $i$ times, thus its Bohr-Sommerfeld
quantization was given as $J_i=i(l_i+const.)$ where $l_i=0,1,2,\cdot \cdot
\cdot $ . This is the same spectrum that would be obtained from string
oscillators, by identifying $J_i=\alpha _{-i}\alpha _i.$ For a closed string
the same arguments lead to the {\it semi-classical} mass spectrum
\begin{equation}
\label{closed}\alpha ^{\prime }(M^2-m_0^2)=\sum_{n=1}^\infty \alpha
_{-n}\alpha _n+\sum_{n=1}^\infty \widetilde{\alpha }_{-n}\widetilde{\alpha }
_n\ .
\end{equation}

The exact quantum spectrum of \ref{ham0} or \ref{ham+} is not known for any
value of $Z$ . But it is known that, for the zero fold sector $Z=0,$ \ref
{ham+} coincides with the 'tHooft equation, and that in the limit of $
m_i^2=0 $ there is an exact zero mass eigenstate ($M^2=0)$ that may be
interpreted as a pion constructed from massless quarks in the chiral
symmetry limit. This interpretation is more appropriate for the string
theory version of the previous section (which gives the same equation) since
it includes fermionic degrees of freedom, thus introducing the concept of
chiral symmetry in the context of string theory. Also, numerical studies of
the spectrum of the 'tHooft equation with massive quarks shows that its
spectrum becomes linear after the first few excited states, thus coming into
agreement with the semi-classical massless string spectrum described by
oscillators that give linearly rising Regge trajectories.

One may also carry out a covariant quantization of the D=2 open or closed
massless string by applying the Virasoro constraints on the Fock space
constructed from {\it two} covariant oscillators $\alpha _n^\mu ,\ \mu =0,1$
. This was done a long time ago \cite{IBa}, and here we give the main
results. By keeping $c=d=2${\it \ } (rather than $c=26$)
{\footnote{The spectrum is dramatically different if c=26. The positive norm
states
become then zero norm states.}}, and taking an intercept $\alpha _0=\frac{
d-2 }{24}=0$ (that is $L_0=0$), it is seen that the only states that satisfy
the Virasoro constraints, $L_n|\psi >=0,\ n\geq 0$ , have positive or zero
norm. The positive norm states are in one-to-one correspondance with the
states described by \ref{closed}, and have the same spectrum and same
degeneracy as the one longitudinal oscillator of \ref{closed}, provided $
m_0^2=0$. The vacuum state $|p^\mu >$ is a massless ``tachyon'' that may be
interpreted as the ``pion'' of the zero-mass 'tHooft equation. Thus, in an
interacting theory (i.e. the $1/N$ corrections) this ``tachyon'' or ``pion''
must decouple (suggestive of Adler zeroes for pions) since massless
particles cannot exist in two dimensions due to their infrared behaviour.
The absence of the ``tachyon'' in the interacting theory has also been
deduced from a very different point of view in the recent work of Gross \cite
{Gross,GoTa}.

\section{Correspondence to recent developments}

In the previous sections I described strings propagating in flat spacetime
and exhibited an action that reproduces 2D QCD interacting with quarks in
flat spacetime. How does this relate to the recent studies of QCD that are
conducted without quarks and in curved spacetime with non-trivial topology?
There are at least two important observations that have been rediscovered
from a different point of view in the recent developments : (i) The
string-QCD correspondance seems to require strings without folds, (ii) The
interacting string, including the $1/N$ correction, has no massless
``tachyon'' or ``pion'' (i.e. even if it is there in the free spectrum, it
must decouple in the interacting theory, or the theory must undergo a phase
transition).

Is there a role that the folds may play to make some further connections?
Note that the mathematical formalism of ``folds'' that we have outlined may
describe different physical situations. A fold may occur by (1) the string
folding back on itself in the opposite orientation, thus making a worldsheet
that folds back on itself, or by (2) wrapping around, which makes a
worldsheet with non-trivial topology. For a folded string lying in one
dimension at a constant time one cannot tell the difference between the two
cases. Therefore, the folded string formalism that we have discussed may
apply to either possibility, however they will have different physical
interpretations from the point of view of color flux tubes. The first
possibility would correspond to having a color charge at the point of the
fold, so that at constant time the color flux goes in and out along the same
line in opposite orientations (one color in, and another color or the same
color out). The second possibility would correspond to the same flux line
(same color) continuing to wrap around until it finds a color charge or its
own starting point. The second possibility can produce a big color flux
passing through the same point, while the first possibility cannot do that.
The QCD-string correspondance tested so far excludes the first kind of fold,
while allowing the second kind to occur only with non-trivial geometrical
backgrounds. This is understandable and expected since the theories
considered do not have any color charges in the adjoint representation of
the color group (no dynamical gluons).

Thus, to test for either kind of folds in the QCD-string correspondance we
can consider enlarging our present 2D theories in two possible directions:
(1) Consider 2D QCD coupled to adjoint matter in either flat or non-trivial
geometries. One example is to consider a sector of 4D QCD in the hybrid
formulation given in \cite{BaPi}; another example is 2D QCD coupled to
adjoint fermions, with or without quarks in the fundamental representation.
This will introduce color charges that can absorb and emit color flux, thus
allowing a fold to occur. (2) Consider strings propagating in non-trivial
topologies (like the recent work on QCD) and allow the string to wrap around
the topology thus producing ``folds'' at the edges of the geometry. For
example a cylinder which is squashed to lie on the table, that is
representing the worldsheet of a 2D string, has the string wrapped around it
at any fixed time, and the string is forced to fold at the edges of the
squashed cylinder. The cylinder may be multiple sheeted, which would lead to
the interpretation that the string is wrapped around many times, thus giving
strings with many folds. Let me say a bit more about tentative results for
either case.

For the enlarged model of QCD coupled to adjoint matter let us consider
bosonic versus fermionic matter. The string-like 'tHooft equation is easily
derived in a Hamiltonian approach in the lightcone gauge $A_{+}=0.$ Such
equations were derived in \cite{BaPi} for the case of bosons, and in \cite
{Kleb} for the case of fermions. For all cases, the Hamiltonian contains
kinetic terms for the matter degrees of freedom and an interaction term that
consists of a linear potential between the total color densities. In the
large $N$ limit this Hamiltonian produces coupled 'tHooft-like equations for
multi-particle states. Thus the two particle state mixes with the three
particle states, etc. The coupling occurs because one can draw leading
planar graphs that include the propagation of the matter fields in addition
to the glue
{\footnote{For quarks in the fundamental representation alone there are only
2-body bound states, and thus no complicated coupling to many body bound
states in the large N limit}}. One may split these equations into two parts:
the first part may be considered ``zeroth order'' in the sense that it
involves only the wavefunctions with a fixed number of matter particles, and
the second part is the coupling. The zeroth order may be associated with
strings binding the particles, and making worldsheets during their
propagation. Then the coupling may be viewed as string-string interactions
that are {\it not} suppressed by factors of $1/N$
{\footnote{However, for reasons that are not well understood this coupling
seems
to be numerically small \cite{Kleb}. I thank M. Douglas for making me aware of
this paper.}}. The zeroth order term may be compared to a string theory
before interactions. For bosonic adjoint matter the details of the
'tHooft-like equation do not match any known string theory. However, for
fermions the 'tHooft-like equation for $Z$ fermions is identically
reproduced by the folded string Hamiltonian given in \ref{hamm}! This is a
new observation that relates strings and large$\ N$ gauge theory. However,
one must emphasize that the wavefunctions for the $Z$ fermions must be
completely antisymmetric, so that only the antisymmetric solutions of these
equations can be admitted. This additional input that comes from the QCD
formulation is missing in the string formulation of \ref{hamm}. So, there is
a bit more to understand.

One may also study strings in curved spacetime in 2D and compare to QCD. One
possibility is to consider higher Riemann surfaces (e.g. cylinder) that are
squashed to lie on the plane since we are in 2D. A Wilson line that is
wrapped around such a cylinder must fold at the end of it. If this cylinder
changes its radius as a function of time, then the folds appear to oscillate
just like the fold solutions of the string theory given in (\ref{sol},\ref
{exsol}). A cylinder that changes as a function of time is possible in QCD
provided it is coupled to dynamical Gravity in 2D. Therefore, one
possibility for including the folds in the string-QCD correspondance is to
consider the larger theory of QCD plus gravity in 2D and compare it to a
string theory on the same curved manifold. So far 2D QCD has been analyzed
on static 2D manifolds. As another example, it should be fun to couple QCD
with or without quarks to the 2D SL(2,R)/R black hole, and study the
relations to the corresponding string theory. I point out that the 2D black
hole string theory has classical solutions that describe folded strings
moving in the vicinity of the black hole.

It is evident that there is much to do even in two dimensions, but I would
urge those interested in this subject to begin to seriously explore the
possible correspondance between 4D QCD and the string action given in \ref
{action}, as outlined at the end of section 2.

\end{document}